\documentclass[%
 reprint,
superscriptaddress,
 amsmath,amssymb,
 aps,
]{revtex4-2}

\usepackage{graphicx}
\usepackage{dcolumn}
\usepackage{bm}
\usepackage{setspace}
\usepackage{bm}
\usepackage{caption}
\usepackage{subcaption}
\usepackage[linguistics]{forest}
\usepackage{booktabs}
\usepackage{color, colortbl}
\usepackage{amsmath}
\usepackage{physics}
\usepackage{setspace}
\usepackage{soul}
\usepackage{color,soul}
\usepackage{threeparttable}
\usepackage{import}

\begin{document}

\title{A first principles study of the Stark shift effect on the zero-phonon line of the NV center in diamond}

\author{Louis Alaerts}
\affiliation{Thayer School of Engineering, Dartmouth College, Hanover, NH 03755, USA}
\author{Yihuang Xiong}
\affiliation{Thayer School of Engineering, Dartmouth College, Hanover, NH 03755, USA}
\author{Sin\'ead Griffin}
\affiliation{Molecular Foundry, Lawrence Berkeley National Laboratory, Berkeley, CA 94720, USA}
\affiliation{Materials Sciences Division, Lawrence Berkeley National Laboratory, Berkeley, CA 94720, USA}
\author{Geoffroy Hautier}
\affiliation{Thayer School of Engineering, Dartmouth College, Hanover, NH 03755, USA}

\date{\today}

\begin{abstract}
Point defects in semiconductors are attractive candidates for quantum information science applications owing to their ability to act as spin-photon interface or single-photon emitters. However, the coupling between the change of dipole moment upon electronic excitation and stray electric fields in the vicinity of the defect, an effect known as Stark shift, can cause significant spectral diffusion in the emitted photons. In this work, using first principles computations, we revisit the methodology to compute the Stark shift of point defects up to the second order. The approach consists of applying an electric field on a defect in a slab and monitoring the changes in the computed zero-phonon line (i.e., difference in energy between the ground and excited state) obtained from constraining the orbital occupations (constrained-DFT). We study the Stark shift of the negatively charged nitrogen-vacancy (NV) center in diamond using this slab approach. We discuss and compare two approaches to ensure a negatively charged defect in a slab and we show that converged values of the Stark shift measured by the change in dipole moment between the ground and excited states ($\Delta \mu$) can be obtained. We obtain a Stark shift of $\Delta \mu$=2.68D using the semi-local GGA-PBE functional and of $\Delta \mu$=2.23D using the HSE hybrid-functional. These values are in good agreement with experimental results. We also show that modern of theory of polarization can be used on constrained-DFT to obtain Stark shifts in very good agreement with the slab computations.


\end{abstract}

\maketitle


\section{Introduction}
Color centers in semiconductors such as the nitrogen-vacancy (NV) center in diamond or the silicon vacancy in SiC are becoming central to quantum technologies. These point defects can act as single-photon sources or as spin-photon interfaces when they have a spin degree of freedom. Quantum defects can enable the generation and distribution of entangled photons in quantum networks, leading to important applications in quantum information science (QIS) such as quantum communication or computing \cite{Atature2018, Ruf2021, Awschalom2018, Pezzagna2021}. The performance of quantum networks strongly depends on the property of the quantum defect. Ideally, it should exhibit a long spin coherence time, a high transition dipole moment, an appropriate emission wavelength, low losses in the phonon sideband of the photoluminescence, and good optical coherence \cite{Wolfoicz2021}. Recent efforts have begun to identify new defects meeting all these requirements using first principles computations \cite{Xiong2023, Ivanov2023,Davidsson2023CaO,Davidsson2023Diamond,Li2022-carbon,Tsai2022,Bassett2019}. 

Among all these properties optical coherence or spectral diffusion is arguably the most difficult property to engineer around \cite{Udvarhelyi2019, Anderson2019}. Spectral diffusion encompasses the variation of the defect emission and of its zero phonon-line (ZPL) with time and space. These variations reduce the degree of indistinguishability of the emitted photons significantly, limiting quantum applications. One of the reason for the shift of ZPL is the Stark shift experienced by a defect. The Stark shift is the change in ZPL wavelength in the presence of an applied electric field. When quantum defects experience electrostatic fluctuations due to other charged defects, their Stark shift modulates the ZPL and leads to spectral diffusion. For most defects, the Stark shift is mainly linear in electric field and originates from the dipole moment change between the defect's ground-state (GS) and excited-state (ES). Therefore, centrosymmetric defects are promising candidates for QIS applications because they are symmetrically protected against a linear Stark shift and should exhibit limited spectral diffusion. In diamond, group IV-vacancy centers such as the silicon-divacancy or the tin-divacancy  are known examples of symmetry-protected defects with good spectral stability \cite{DeSantis2021, Aghaeimeibodi2021, Rogers2014, Sipahigil2014}. More recently, defects in new hosts such as CaWO$_4$ have shown very limited spectral diffusion due to their symmetry as well \cite{Ourari2023}. We note that important hosts lacking centrosymmetry such as most polymorphs of silicon carbide do not have such a symmetry-forbidden linear Stark shift.

First-principles computations offer a powerful way to compute defect properties to aid in understanding and designing new color centers for QIS \cite{Dreyer2018, Gali2023}. Two approaches have been used to compute the Stark shift using density functional theory (DFT). The first approach monitors the ZPL changes as a function of the applied electric field \cite{Bathen2020} while the second consists of directly computing the dipole moment of the GS and the ES using the modern theory of polarization \cite{Udvarhelyi2019}. Applying electric fields in systems with periodic boundary conditions requires the use of slab geometries where the electric field can be terminated in the vacuum, resulting in computationally expensive calculations. On the other hand, the modern theory of polarization can be used directly in a periodic supercell - this is appealing since supercells are the primary way to model defects with first principles. However, this approach suffers from several drawbacks. Firstly, only the linear component of the Stark shift is accessible. Secondly, absolute values of polarization are ill-defined in periodic systems, only polarization changes ($\Delta P$) are physically meaningful \cite{Spaldin2012}. In practice, $\Delta P$ is defined with respect to a centrosymmetric reference which intrinsically has no polarization. In the case of ferroelectric materials for which the modern theory of polarization is often used, the centrosymmetric and polar structures marginally differ because, by definition, polarization switching should be possible and occur through a centrosymmetric reference. Finding this reference is therefore typically trivial. On the other hand, for points defects, there is no such guarantee, especially when the host itself is polar (\textit{e.g.} SiC, ZnO). Next, the polarization is only defined for insulating states. This can be challenging for point defects that introduces shallow levels in the band gap as the structural deformation needed to form the centrosymmetric reference state could shift these levels into the conduction band and create metallic-like states which preclude the use of the modern theory of polarization. Finally, DFT calculations on ES are often carried out using the $\Delta$-SCF method which consists of imposing the desired occupation of the electronic states. It is not clear if this constrained-DFT approach can be used seamlessly in combination with the modern theory of polarization For all these reasons, alternative methods for the calculation of the ZPL energy as a function of the applied electric field are needed. 

In this study, we investigate the Stark shift of the negatively charged NV center in diamond using a slab model by monitoring the energy change between the GS and the ES under different values of applied electric field. We focus on the NV center because it is an extremely well-studied quantum defect and its Stark shift has already been studied both experimentally and theoretically. We show that slabs can be used to obtain a converged value for the Stark shift and the change in dipole moment between GS and ES. We also perform calculations within the modern theory of polarization formalism and demonstrate that both approach gives excellent agreement when electronic occupation is carefully imposed in the excited state. We discuss this change in dipole moment and compare it to experiments and previous theoretical results using the modern theory of polarization. Our work not only provides important information on the NV center in diamond but an analysis of the methodological challenges in computing Stark shifts for any defect using slabs or the modern theory of polarization.

\section{\label{sec:level1} Results}

\subsection{Developing a slab model for charged defect}

\begin{figure}
        \centering
        \includegraphics[width=0.45\textwidth]{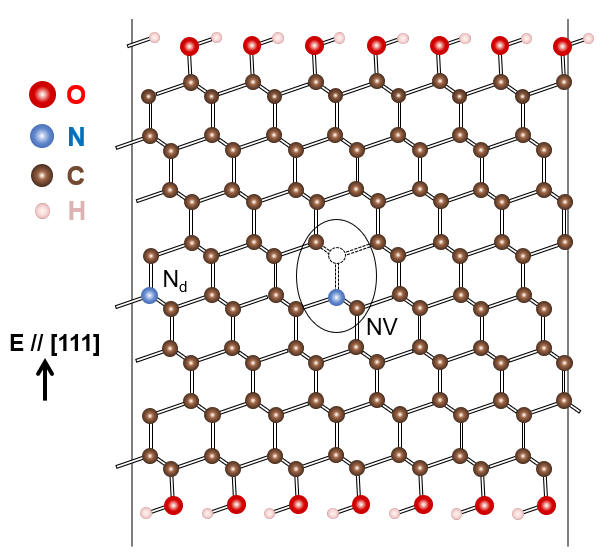}
        \caption{Sketch of a slab used to calculate Stark shift showing the hydroxyl termination as well as the NV center (in the center) and the N dopant atom (on the left). This slab is about 15\AA\ thick (excluding the -OH termination) and the distance between the dopant and the NV center is about 9.11\AA.
        \label{fig:slab_model}}
\end{figure}

The Stark shift of the ZPL of a defect can be computed from the change of energy in both the ground and excited state under an electric field. The most common approach to model point defects is to use a large supercell and periodic boundary conditions. However, since applying an electric field on a periodic system is methodologically challenging, we choose to work with a slab configuration in which we position the defect (see Fig.\ref{fig:slab_model}). We built a (111)-oriented diamond slab terminated by hydroxyl groups (-OH) to avoid dangling bonds \cite{Chou2017}. The NV center is positioned in the center of the slab, with its nitrogen-to-vacancy axis along the (111) direction, also in the direction of our applied field. The structure of the defect was relaxed once placed inside the slab.

While using a slab allows to break the periodicity in order to apply an electric field easily within periodic boundary conditions, it also leads to another difficulty concerning the charge state of the NV center. In supercells, charged defects are modelled by artificially adding (subtracting) an electron and compensating the charge imbalance by a background positive (negative) charge through a jellium model. This approach is challenging in slabs because spurious electrostatic interactions in the vacuum region lead to a nonphysical dependence of the total energy on the vacuum size \cite{Komsa2014}. Fortunately, there are at least two methods to circumvent this problem. One consists in adding a dopant atom inside the slab \cite{Kaviani2014}, preferably in the same layer as the defect to minimize the dipole-dipole interaction between the dopant and the defect \cite{Lofgren2018,Gali2019}. A sketch of this slab showing the hydroxyl termination, the NV, and the dopant lying in the same layer is illustrated on Fig.\ref{fig:slab_model}. We chose here a nitrogen substitution of carbon as a dopant that will donate electrons to the NV center, making it negatively charged. The other approach we considered was inspired by the work from Richter et al. \cite{Richter2013}. In this work on MgO surfaces, the authors slightly change the atomic number of all Mg atoms by $\frac{-q}{N_{Mg}}$, where \textit{q} is the targeted charge state, to mimic doping while keeping the whole system neutral and avoiding issues with charged slabs. However, directly changing the atomic number in the DFT code used in this work is difficult because it relies on pseudopotentials. We circumvented this by mixing all carbon atoms with a minute amount of nitrogen ($\approx$ 0.03 \%) using the virtual crystal approximation (VCA) method \cite{VCA}. For the sake of conciseness, we will focus on the dopant-containing slab in the rest of the paper since we find that the conclusions are essentially the same for the two doping schemes.

Calculations with slabs and of dilute defects require careful convergence. We first perform a convergence study at the GGA-PBE level on two parameters: the thickness of the slab, and the dopant-NV distance, focusing on the electronic structure of the defect and its calculated Stark shift. The detailed results of this study can be found in the Section 1 of the Supplementary Material. We observed that a thickness of approximately 15\AA\ (excluding the hydroxyl terminations) and a dopant-NV distance of 9.11\AA\ lead to an acceptable convergence. This corresponds to a simulation box of 979 atoms. All of the following results are reported for this slab. We also confirmed that the dopant had no impact on the calculated Stark shift by switching N with P and observed no difference (see Supplementary Material, Section 1.3). 

\subsection{Comparison between the electronic structure and ZPL for the slab and bulk}

We begin by demonstrating that our slab captures the electronic structure of the NV center by comparing the Kohn-Sham single-particle levels and the charge density contour plots (see  Supplementary Material, Section 2) to bulk calculations. We also compare the ZPL and the vertical transition energy (defined as the electronic transition from $^3A_2$ to $^3E$ at fixed geometry) (see Table \ref{table:electronic_properties_NV}) to bulk calculations and experimental data.

\begin{figure}
        \centering
        \includegraphics[width=0.45\textwidth]{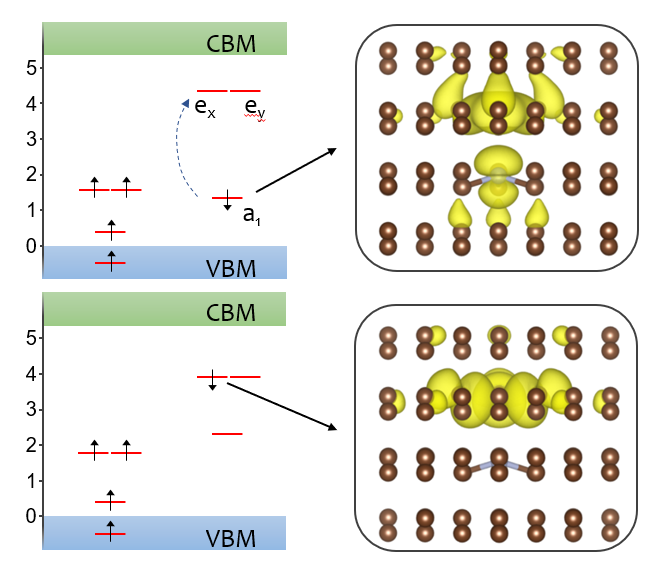}
        \caption{Calculated Kohn-Sham eigenvalues and corresponding charge density of the NV center using a bulk model showing the ZPL electronic transition (dashed arrow)}
        \label{fig:ks_level}
\end{figure}

The experimental ZPL optical transition lies at 1.945eV and corresponds to a spin-conserving transition from the triplet GS ($^3A_2$) to the ES ($^3E$) \cite{Davies1976, Doherty2012}. In terms of Kohn-Sham single particle orbitals, this corresponds to an electronic transition from the $a_1$ to the degenerate $e_{x}, e_{y}$ states, resulting in a significant change in the electronic density around the defect (see Fig. \ref{fig:ks_level}) and a change of dipole moment. The excited state is obtained by constrained-DFT where the unoccupied single-particle energies (here, $e_{x}$ and $e_{y}$) are constrained to be occupied. To model the dynamic Jahn-Teller effect present in the NV, we impose the occupation of $e_{x}$ and $e_{y}$ with half an electron each. This effectively imposes the C$_{3v}$ symmetry of the ground state for the excited state as well \cite{Thiering2017}.

The calculated values for the slab and the bulk calculations are in good agreement but both underestimate the experimental transition energies $\Delta E_{ZPL}$ (see Table \ref{table:electronic_properties_NV}). This is a well-known issue of the GGA-PBE functional \cite{Cohen2008}. Typically, hybrid functionals are used to obtain better estimations of the band gap \cite{Chen2012}. Here, we used the Heyd-Scuseria-Erzerhof (HSE) \cite{Heyd2003}, which contains 25\% of the exact exchange, finding the calculated ZPL and vertical transition energies to be 2.017eV and 2.372eV, which compare well with our own bulk HSE calculations as well as with the experimental results.

As previously described \cite{Lofgren2018}, the presence of a dopant atom leads to a small splitting of the degenerate $e_x$ and $e_y$ levels. In our slabs, we observe that this energy splitting is about 20meV and that even at larger dopant-NV distance the splitting remains above 10meV. Consistent with this picture, our VCA calculations, for which there are no dopant-NV interactions, show a splitting of less than 0.5meV (see Supplementary Material, Section 1.1). Our convergence study of the dopant-NV distances shows that this effect does not affect the calculated Stark shift (see Supplementary Material, section 1.2). 

\begin{table}[h!]
    \centering
    \begin{tabular}{lcc}
    \toprule
    Method &  $\Delta E_{ZPL}$ (eV)   &  $\Delta E_{e_x - e_y}$ (meV) \\
    \midrule
    Doped-slab (PBE)                              &   1.720         & -19.6                    \\
    Bulk (PBE)                                &   1.687         & 0                        \\
    Doped-bulk (PBE) \cite{Lofgren2018}       &   1.701         & 66\tnote{*}$^\ast$       \\
    \midrule
    Doped-slab (HSE)                                &   2.017         & -23.2                    \\
    Bulk (HSE)                                &   1.997         & 0                        \\
    \midrule    
    Exp \cite{Davies1976}                   &   1.945         & N.A                      \\
    \bottomrule
    \end{tabular}
    \begin{tablenotes} \footnotesize
    \item[*] $^\ast$  The splitting reported here is actually the energy difference between the excited-state with the electron in the $e_x$ state and the excited-state with the electron in the $e_y$ state
    \end{tablenotes}
    \caption{Comparison of the dopant-slab transition energies with their bulk and experimental counterparts. We also compare our bulk calculations, where the negative charge is artificially added, with previously published bulk calculation which uses a dopant atom.}
    \label{table:electronic_properties_NV}
\end{table}

\subsection{Stark shift first principles computations}
Before we can study the Stark shift using our slab model, we have to find the relationship between the field applied on the full system (slab and vaccuum) and the electric field present inside the slab. This is done by a linear fitting of the averaged electrostatic potential inside a pristine diamond slab terminated with the same hydroxyl groups as the NV-containing slab for different values of applied electric field. We find that this local electric field is simply screened by a factor of 5.82 (see Supplementary Material, section 3.), which is close to the dielectric constant of diamond ($\approx$ 5.7). 

The two quantities we are aiming to evaluate are the dipole moment change $\Delta \mu$ and the polarizability change $\Delta \alpha$ which corresponds to the linear and quadratic coefficients, respectively, of the expansion of the energy change $\Delta E_{ZPL}$ between the GS and the ES with respect to the macroscopic electric field $F$ (see Eq. \ref{eq:energy_field_expansion}). The dipole moment change is a vector and the polarizability change is a second-order tensor. Here, because of the geometry of the defect within the slab, we are only probing $\Delta \mu_z$ and $\Delta \alpha_{zz}$, their components along the main axis of the defect (i.e., z is along the nitrogen to vacancy axis).

\begin{equation}
        \label{eq:energy_field_expansion}
        \Delta E_{ZPL} = - \Delta \mu_{z} F_{z} - \frac{1}{2} \Delta \alpha_{zz} F_{z}^2
\end{equation}

\begin{figure}
    \centering
    \includegraphics[width=0.45\textwidth]{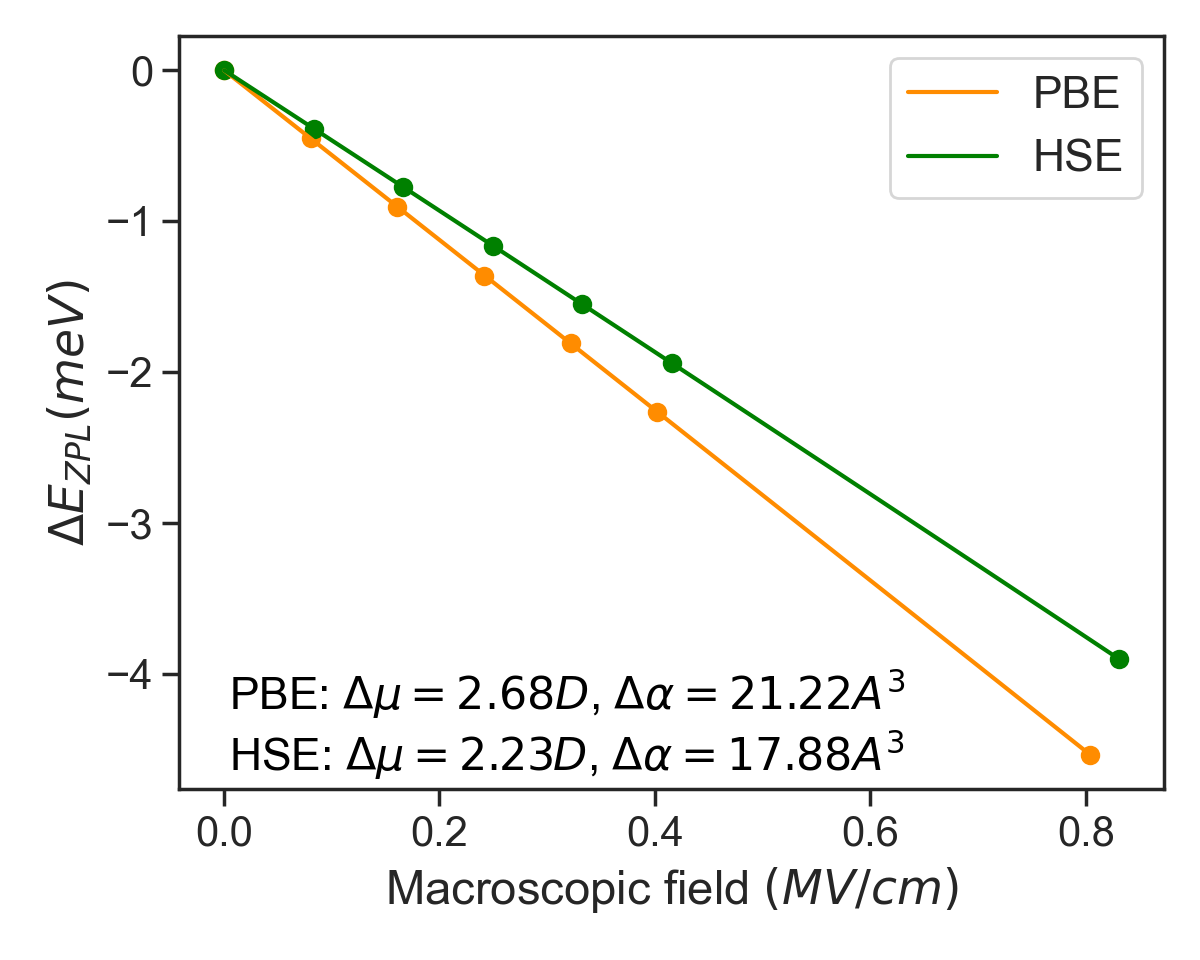}
    \caption{Calculated Stark shift of the $^3A_2$ to $^3E$ electronic transition using the semi-local exchange-functional GGA-PBE and hybrid exchange-functional HSE.}
    \label{fig:stark_shift}
\end{figure}

To calculated this, we consider the total energy of the $^3A_2$ GS and of the $^3E$ ES for different values of electric field and fit the set of points to Eq. \ref{eq:energy_field_expansion}. During this step, the ions were kept fixed. Fig.\ref{fig:stark_shift} plots how the ZPL energy changes with the field in the slab. We find a linear dependence, highlighting that the response primarily originates from the change of dipole moment. For the converged slab (thickness $
\approx$ 15\AA\ and dopant-NV distance of 9.11\AA), we obtain $\Delta \mu_z$ = 2.68D and $\Delta \alpha_{zz}$ = 21.22\AA$^3$ (in units of polarizability volume: $\alpha / 4\pi\epsilon_0$) using GGA-PBE. We repeated this methodology for other thicknesses and dopant-NV distances and found that $\Delta \mu_z$ was never below 2.37D but never reaches 2.71D for the tested geometries while $\Delta \alpha_{zz}$ varied between 19.03\AA$^3$ and 26.43\AA$^3$. We note that the convergence is rather erratic (see Fig. 1 and 2 of the Supplementary Material) but considering the large number of calculations performed, we are confident that the calculated $\Delta \mu$ is approximately 2.5D. Furthermore, using the alternative doping method (VCA), our calculations quickly converged to give $\Delta \mu_z$ = 2.18D and $\Delta \alpha_{zz}$ = 22.14\AA$^3$ (see Fig. 3 of the Supplementary Material). 
Interestingly, the choice of functional seems to only have a limited effect on the calculated values. The slab size showing convergence at the GGA-PBE level gives values of $\Delta \mu_z$ = 2.23D and $\Delta \alpha_{zz}$ = 17.88\AA$^3$ using HSE (see Fig. \ref{fig:stark_shift}). We also performed calculations at higher electric field for the PBE slab without observing any significant deviation on the calculated coefficients (see Fig. 5 of Supplementary Material).

Differences in single-particle energy between occupied and unoccupied states are often used as a proxy for the vertical excitation or even ZPL. We can also evaluate the Stark shift by directly using the Kohn-Sham energies of the $a_1$ and $e_x$ levels, obtained from our calculations on the $^3A_2$ GS. This method gave values of 4.24D for $\Delta \mu$ and 132.25\AA$^3$ using GGA-PBE and 2.95 and 232.65\AA$^3$ using HSE (see Supplementary Material, Section 4), indicating that evaluating $\Delta \mu$ from the Kohn-Sham energies is not reliable . 

\subsection{Modern theory of polarization calculation}
The modern theory of polarization can be used to obtain directly the polarization of a supercell and thus dipole moment of the defect. This approach does not require a slab and should provide comparable values to our slab computations. However, previous work reported a change of dipole moment ($\Delta \mu$ = 4.43D), significantly larger than the dipole moment calculated within the slab approach ($\Delta \mu$ = 2.68D) presented here \cite{Udvarhelyi2019}. We believe this discrepancy comes from an implementation issue within the VASP code when imposing the occupation of the e$_x$ and e$_y$
orbitals to model the excited state on top of the modern theory of polarization routine. Our own modern theory of polarization computations using VASP are in good agreement with the previous results (4.34D vs 4.43D). However, a careful inspection of the code indicates that the electron occupation is not imposed as expected. We have implemented a methodology to correct the electron occupation and performed a modern theory of polarization computations (see Supplementary Material, Section 5 for details). We have used a carefully chosen centrosymmetric reference to perform these computations as the polarization is defined up to a quantum of polarization within this theoretical framework. We obtain a Stark shift value of 2.82D using GGA-PBE and 2.23D using HSE in excellent agreement with our slab computations.

\section{Discussion}
The calculated dipole moment changes are in good agreement with previous experimental results, with reported values of 1.4D for NV ensembles, and ranging from 1.5D and 2.82D for individual NV centers \cite{Block2021, Tamarat2006, Ji2024}. On the other hand, our calculated polarizability changes not only underestimate the experimental value by 3 orders of magnitude ($\Delta \alpha \approx -6\times10^4$\AA$^3$) but also predict an increase, \textit{i.e.} the ES is more polarizable, instead of a decrease. Because the polarizability is a second-order tensor, a possible explanation for the discrepancy between the experimental changes of polarizability and the calculated ones is that we are comparing different components. In our calculations, the NV is parallel to the direction of the field and we only probe $\alpha_{zz}$ (see Eq. \ref{eq:energy_field_expansion}). In experiments, the orientation of the defect is not known so different components of the tensor cannot be distinguish. This is reflected by the wide spread of reported values (between 0 and -6$\times10^4$ \AA$^3$). Overall, reported polarizability changes can strongly vary from a system to another. While typical values in molecules are positive and on the order of tens of \AA$^3$ \cite{Brunel1999, Schadler2019}, in other defects such as the blue color center in \textit{h}-BN, measurements on two different emitters yielded values of -168\AA$^3$ and 1078\AA$^3$ \cite{Zhigulin2023}. 

A slab model was previously used to compute the Stark shift of a point defect \cite{Bathen2020}. Applying this methodology to the Si vacancy in SiC, Bathen et al. calculated a dipole moment change of 1.46D, about one order of magnitude above the calculated dipole moment using the modern theory of polarization (0.21D) \cite{Udvarhelyi2019}. Experimental measurements are also scattered, with reported values ranging between 0.18D \cite{Ruf2021} and 0.72D \cite{Lukin2020}, making a comparison with theory difficult. However, in both cases, the slab model overestimates the linear response of the Stark shift with possible sources of errors being the thickness of the slab, the vacuum size or the exchange-correlation functional (GGA-PBE) \cite{Bathen2020}. While every system would require different convergence parameters, we note that the slabs used in Bathen et al. are smaller (250 atoms, excluding terminations) than the needed slabs for obtaining converged results in our study (399 atoms for VCA, 575 atoms using a dopant). This supports the possibility that the discrepancy of the reported dipole moments between the slab calculation of \cite{Bathen2020} and the experimental results of \cite{Udvarhelyi2019} and \cite{Lukin2020} is due to size effects.

To the best of our knowledge, three studies have tried to calculate the magnitude of the ZPL transition Stark shift in the NV center. Maze \textit{et al.} used a molecular cluster of about 71 atoms and terminated by hydrogen atoms on which they applied an electric field but only semi-quantitative results ($\Delta \mu$ = 0.79D) were obtained due to the small size of the molecular cluster \cite{Maze2011}. Another study, based on a supercell approach combined with the modern theory of polarization found the calculated dipole moment $\Delta \mu$ to be 4.34D \cite{Udvarhelyi2019}. Based on our own computations within the modern theory of polarization using the same code VASP, this value is however questionable and we obtained a value of 2.82D to 2.23D within GGA-PBE and HSE respectively. The agreement between our results using the modern theory of polarization and slabs reinforce their validity.
Finally, the Stark shift of the NV center has recently been computed using a beyond-DFT approach and a quantum embedding approach in which the dipole moment change is directly evaluated from the many-body wave function built from a Wannierization procedure. They obtained $\Delta \mu$ = 1.6D \cite{Lopez2024}. We note that the calculation of dipole momenta using Wannier functions is formally equivalent to the use of the modern theory of polarization but that their approach includes correlation effects that are absent in DFT \cite{Spaldin2012}.

\section{Conclusion}
In summary, using DFT calculations, we demonstrate that the Stark shift of the NV center in diamond, can be quantitatively calculated by applying an electric field to a defect in a slab. To address the challenge of using a charged defect in a slab, we insert an additional nitrogen atom in the same atomic layer to act as a dopant, inside a (111)-oriented diamond slab terminated with hydroxyl groups. We used two different functionals, GGA-PBE and HSE, and showed that in both cases, the electronic properties are in good agreement with supercell calculations. Our calculations show that the Stark shift of the NV center is dominated by the dipole moment change, with values of 2.68D and 2.23D using the GGA-PBE and HSE functional, respectively, in good agreement with previous experimental results. Using the modern theory of polarization, we also obtained value in good agreement with these slab values. This indicates that modern theory of polarization can be used to compute the Stark shift on defects provided electron occupation is well constrained to model the excited state.

While we focused here on the NV center, our approaches can be readily applied to other quantum defects and provide important insight into their ZPL Stark shift and spectral diffusion. 

\section{Methodology}
All calculations were carried out with the Vienna ab-initio simulation package (VASP), a density functional theory (DFT) plane-wave code \cite{DFT1, VASP1, VASP2} based on the projected augmented wave method formalism \cite{PAW1}. The exchange-correlation potential was described using the generalized gradient approximation (GGA) of Perdew-Burke-Ernzerhof (PBE) \cite{PBE}. We also used the Heyd-Scuseria-Ernzerhof (HSE) function with 25\% of exact exchange. Except specified otherwise, all the calculations were performed using a $\Gamma$-point only grid with the plane-waves basis energy cutoff set to 400eV. 

For the GGA-PBE bulk calculations, the NV center was placed inside a 4x4x4 supercell with its principal axis oriented along the (111)-direction. The supercell size was set to 3x3x3 for the HSE calculations. The electronic self-consistent loop was converged down to $10^{-6}$eV and the ionic relaxation was stopped once the forces on the atoms were smaller than 0.01eV/\AA. For the slab calculations, we used the same electronic self-consistent loop and ionic relaxation criteria as for the bulk. The relaxation was done in two steps. First, the hydroxyl termination (see main text) and the first two layers of carbon were relaxed, then we fixed the hydroxyl termination and we relaxed all the internal layers of the slab. All the ions were kept fixed during the electric field calculations. The NV center was placed in the middle of a (111)-oriented diamond slab terminated by hydroxyl groups, with its principal axis parallel to the direction of the applied electric field. The convergence of the electronic properties and of the Stark shift with respect to the slab thickness was carefully checked (Supplementary Material, Section 1.2). We found out that a thickness of about 14.95\AA\ (18.43\AA\ including the hydroxyl termination) was enough. Periodic images of the slabs were separated by a vacuum region of 22\AA\ to avoid spurious interactions. We checked the convergence of the electronic properties and of the Stark shift with respect to (1) the distance between the NV center and the dopant atom in the case of the doped slab and (2) the dilution factor of the nitrogen in the case of the VCA-slab (Supplementary material, Section 1.2). In both cases, this was done by varying the in-plane lattice parameters of our slab.

We used the $\Delta$-SCF method to constraint the occupation of states for the excited state. In order to remove the effect of dynamical Jahn-Teller distortions, the $e_x$ and $e_y$ orbitals were occupied equally, with an half electron, in the excited-state \cite{Thiering2017}. More details about the calculation of the dipole moment using the modern theory of polarization can be found in the Supplementary Material. Pymatgen \cite{Ong2013} was used to prepare the calculation inputs as well as to process and analyze the results.

\begin{acknowledgments}
We thank Alp Sipahigil, Chris P. Anderson and Wei Chen for useful discussions. 
This work was supported by the U.S. Department of Energy, Office of Science, Basic Energy Sciences 
in Quantum Information Science under Award Number DE-SC0022289. 
This research used resources of the National Energy Research Scientific Computing Center, 
a DOE Office of Science User Facility supported by the Office of Science of the U.S.\ Department of Energy 
under Contract No.\ DE-AC02-05CH11231 using NERSC award BES-ERCAP0020966. 
\end{acknowledgments}

\bibliography{biblio}

\end{document}


\maketitle

\section{Convergence studies}

\subsection{Electronic transition energies}

We checked the convergence of two important electronic transition energies, $\Delta E_{ZPL}$ and $\Delta E_{vertical}$. The first is defined as the energy difference between the ground-state $^3A_2$ and the excited-state $^3E$, each in their own relaxed geometry,  while the latter is defined as the energy difference between the ground-state and the excited-state, both in the ground-state geometry. We also monitor the splitting caused by the dopant atom between the degenerate level $e_x$ and $e_y$ \cite{Lofgren2018}. First, we vary the thickness of the slab while keeping dopant-NV distance constant d$_{Nd-NV}$=9.11\AA  (see Table \ref{table:electronic_properties_thickness_convergence}). Then, we study the convergence with respect to the distance between the dopant atom and the NV center at a constant slab thickness of t=14.95\AA, excluding the hydroxyl termination (see Table \ref{table:electronic_properties_in-plane_convergence}). For the VCA slab, we checked the side-length convergence using the same thickness as for the dopant-NV distance convergence study (see Table \ref{table:electronic_properties_VCA_convergence}). Experimental and bulk calculations values are given for reference.

\begin{table}[H]
\begin{tabular}{cccc}
\toprule
Thickness (\AA) & $\Delta E_{ZPL}$ (eV)   & $\Delta E_{vertical}$ ($a_1$ to $e_x$) &  Splitting ($e_x - e_y$) (meV) \\
\midrule
14.95                                   & 1.720    & 1.889             & -19.6           \\
19.56                                   & 1.692    & 1.886             & -39.7           \\
27.34                                   & 1.704    & 1.895             & -19.5           \\
29.40                                   & 1.723    & 1.898             & -19.2           \\
\midrule
Bulk (nelect)                           &   1.705    & 1.864             & 0             \\
Bulk (dopant) \cite{Lofgren2018}        &   1.701    & 1.887             & 66\tnote{*}$^\ast$            \\
Exp \cite{Davies1976}                   &   1.945    & 2.180             & N.A           \\
\bottomrule
\end{tabular}

\begin{tablenotes} \footnotesize
\item[*] $^\ast$  The splitting reported here is actually the energy difference between the excited-state with the electron in the $e_x$ state and the excited-state with the electron in the $e_y$ state
\end{tablenotes}

\caption{Convergence of the electronic properties with respect to the slab thickness, at a constant dopant-NV distance of 9.11\AA. For reference, the corresponding experimental values as well as for bulk calculations using two different doping schemes (nelect, dopant) were added.}
\label{table:electronic_properties_thickness_convergence}
\end{table}

\begin{table}[H]
\centering
\begin{tabular}{cccc}
\toprule
Distance (\AA) & $\Delta E_{ZPL}$ (eV)  & $\Delta E_{vertical}$ (eV) & Splitting (${e_x - e_y}$) (meV) \\
\midrule
7.58                                    &   1.761    & 1.930             & -21.8                    \\
9.11                                    &   1.720    & 1.889             & -19.6                    \\
10.10                                   &   1.705    & 1.880             & -11.8                    \\
11.58                                   &   1.699    & 1.874             & -11.9                     \\
\midrule
Bulk (nelect)                           &   1.705    & 1.864             & 0                         \\
Bulk (dopant) \cite{Lofgren2018}        &   1.701    & 1.887             & 66\tnote{*}$^\ast$                  \\
Exp \cite{Davies1976}                   &   1.945    & 2.180             & N.A           \\
\bottomrule
\end{tabular}

\caption{Convergence of the electronic properties with respect to the distance between the dopant atom and the NV. The slab thickness was kept constant (t=14.95\AA). For reference, the corresponding experimental values as well as for bulk calculations using two different doping schemes (nelect, dopant) were added.}

\label{table:electronic_properties_in-plane_convergence}
\end{table}

\begin{table}[H]
\begin{tabular}{cccc}
\toprule
Side-length (\AA) & $\Delta E_{ZPL}$ (eV)   & $\Delta E_{vertical}$ ($a_1$ to $e_x$) &  Splitting ($e_x - e_y$) (meV) \\
\midrule
12.63                                   & 1.861    & 2.028             & -0.4          \\
15.16                                   & 1.761    & 1.939             & -0.3          \\
17.69                                   & 1.722    & 1.903             & -0.1          \\
\midrule
Bulk (nelect)                           &   1.705    & 1.864             & 0                         \\
Bulk (dopant) \cite{Lofgren2018}        &   1.701    & 1.887             & 66\tnote{*}$^\ast$                  \\
Exp \cite{Davies1976}                   &   1.945    & 2.180             & N.A           \\
\bottomrule
\end{tabular}
\caption{Convergence of the electronic properties with respect to the side-length and using the VCA method to dope the slab. The slab thickness was kept constant (t=14.95\AA).  For reference, the corresponding experimental values as well as for bulk calculations using two different doping schemes (nelect, dopant) were added.}
\label{table:electronic_properties_VCA_convergence}
\end{table}

\subsection{Stark shift convergence}

\subsubsection{Convergence study of the slab thickness using a dopant atom}
\begin{figure}[H]
    \centering
    \includegraphics[width=1.0\textwidth]{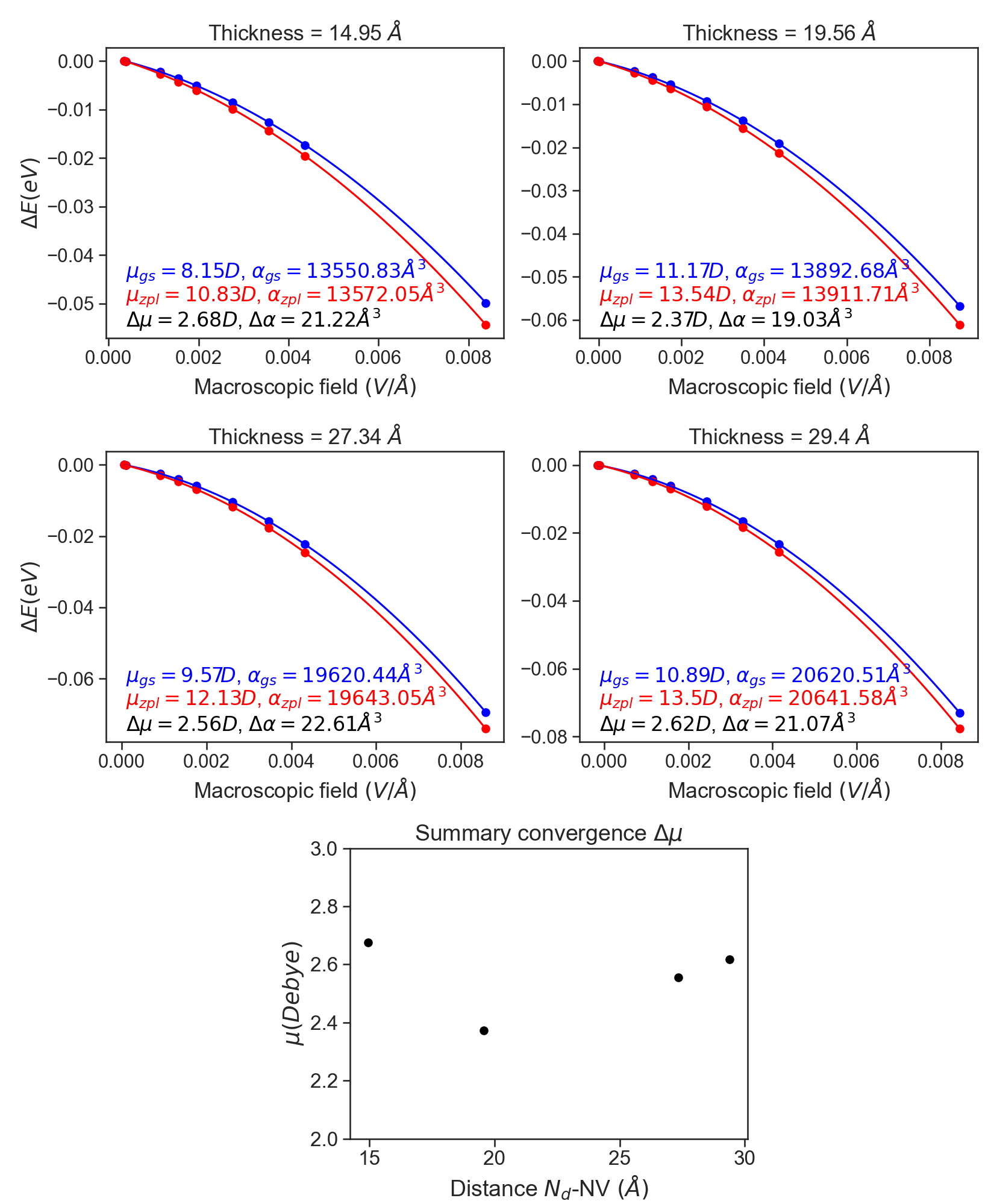}
    \caption{Stark shift convergence of dopant-slabs with respect to the slab thickness. The absolute values of the ground-state (in blue) and excited-state (in red) are pretty sporadic, varying from 8.15 to 11.17D for the ground-state and from 10.83 to 13.54 for the excited state. On the other hand, the change of dipole moment barely varies with the slab thickness, with values around 2.4 and 2.68D.}
    \label{fig:convergence_thickness}
\end{figure}

\subsubsection{Convergence study of the dopant-defect distance using a dopant atom}

\begin{figure}[H]
    \centering
    \includegraphics[width=1.0\textwidth]{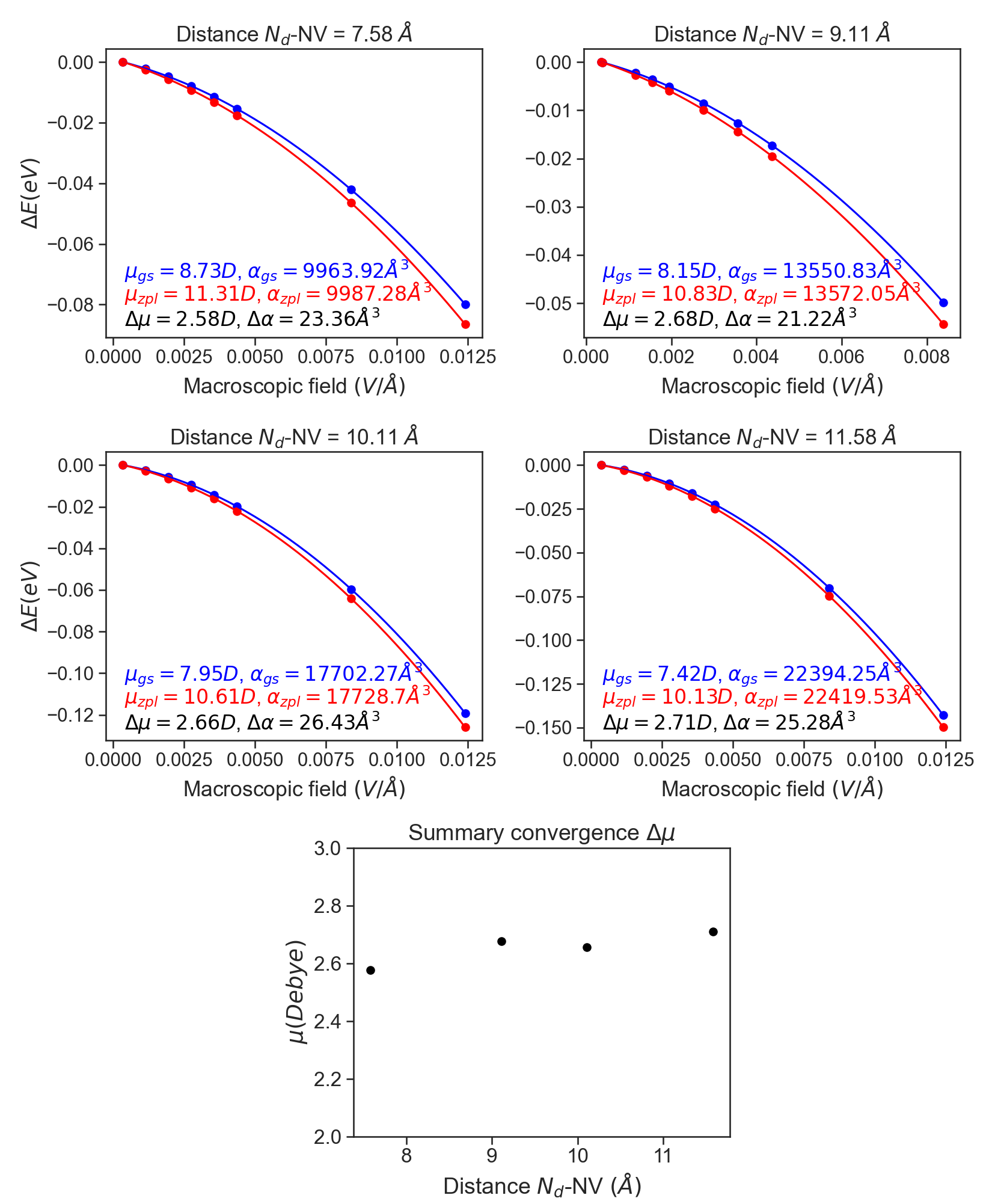}
    \caption{Stark shift convergence of dopant-slabs with respect to the dopant-NV distance. The absolute values of the ground-state (in blue) and excited-state (in red) are steadily diminishing when the distance is increased, decreasing from 8.73 to 7.42D for the ground-state and from 11.31 to 10.13 for the excited state. The change of dipole moment slightly fluctuates with the dopant-NV distance, with values around 2.58 and 2.71D.}
    \label{fig:convergence_in-plane}
\end{figure}

\subsubsection{Convergence study of the side-length distance using the VCA method}

\begin{figure}[H]
    \centering
    \includegraphics[width=1.0\textwidth]{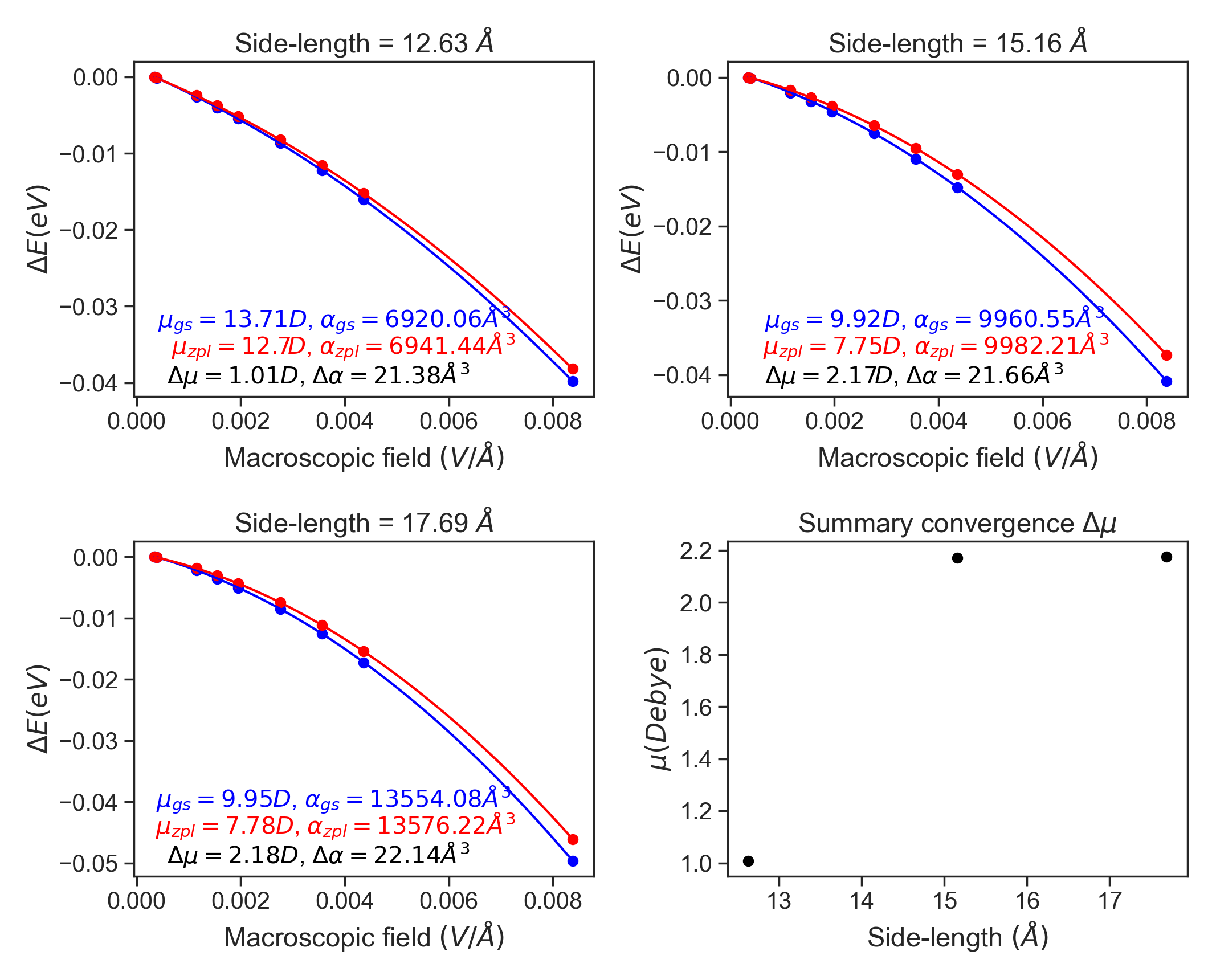}
    \caption{Stark shift convergence of VCA-slabs with respect to the slab side-length. Both the absolute values and the change of dipole moments converge quickly with the side-length.}
    \label{fig:convergence_vca}
\end{figure}

\subsection{Comparison between two dopant, N and P}
To ensure that the doping specie had not impact on the calculated Stark shift, we substitute N for P. The dipole moment change for the N-doped slab is 2.68D and 2.69D with P while the polarizability changed from 21.22\AA$^3$ to 10.24\AA$^3$. Since the Stark shift is dominated by the linear response within the considered applied electric field range, we conclude that the dopant specie has no effect on the result. 

\begin{figure}[H]
    \centering
    \includegraphics[width=1.0\textwidth]{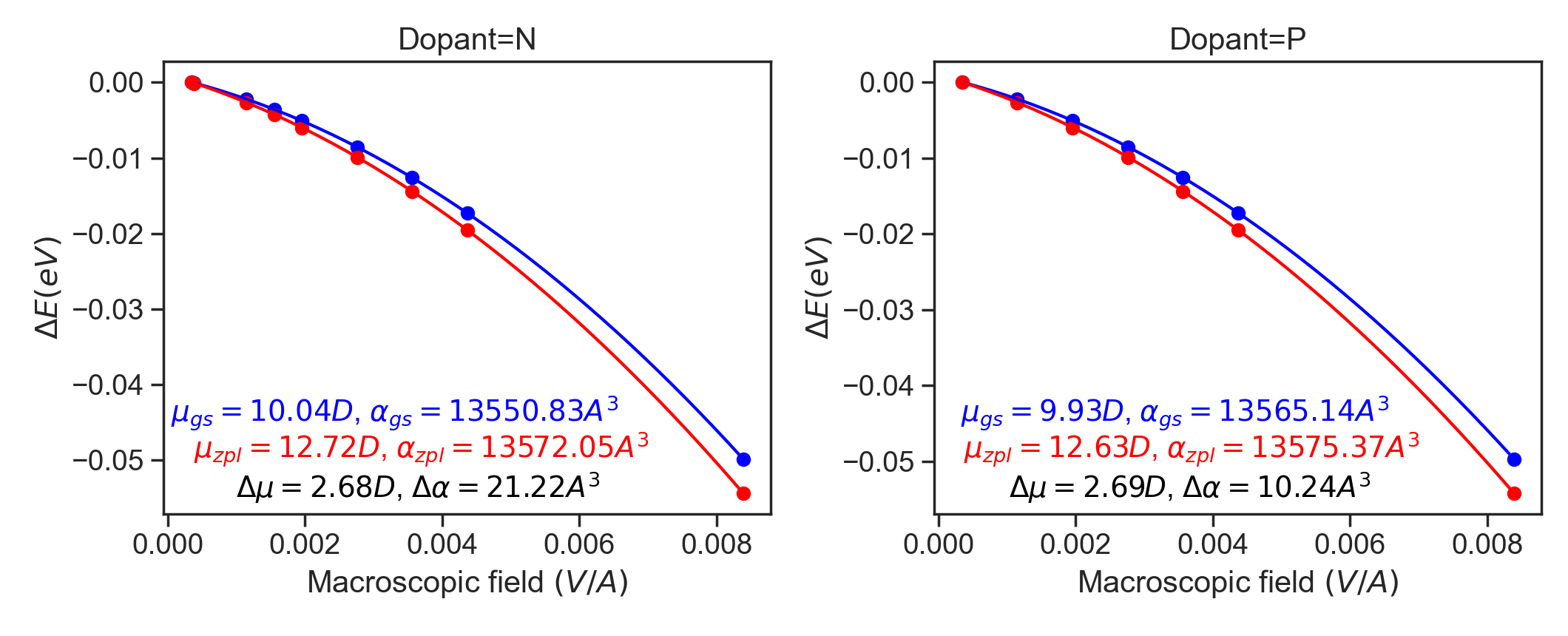}
    \caption{Calculated Stark shift for a 15 \AA\ thick slab with a dopant-defect distance of 9 \AA\ and for the two different dopant: N (left) and P (right) }
    \label{fig:convergence_dopant}
\end{figure}

\subsection{Calculations at higher field}
We performed calculation at higher electric field to make sure that the calculated dipole moment and polarizability change did not depend on the intensity of the applied field. We found that both quantities are almost unaffected by the addition of points at higher electric field. 

\begin{figure}[H]
    \centering
    \includegraphics[width=0.45\textwidth]{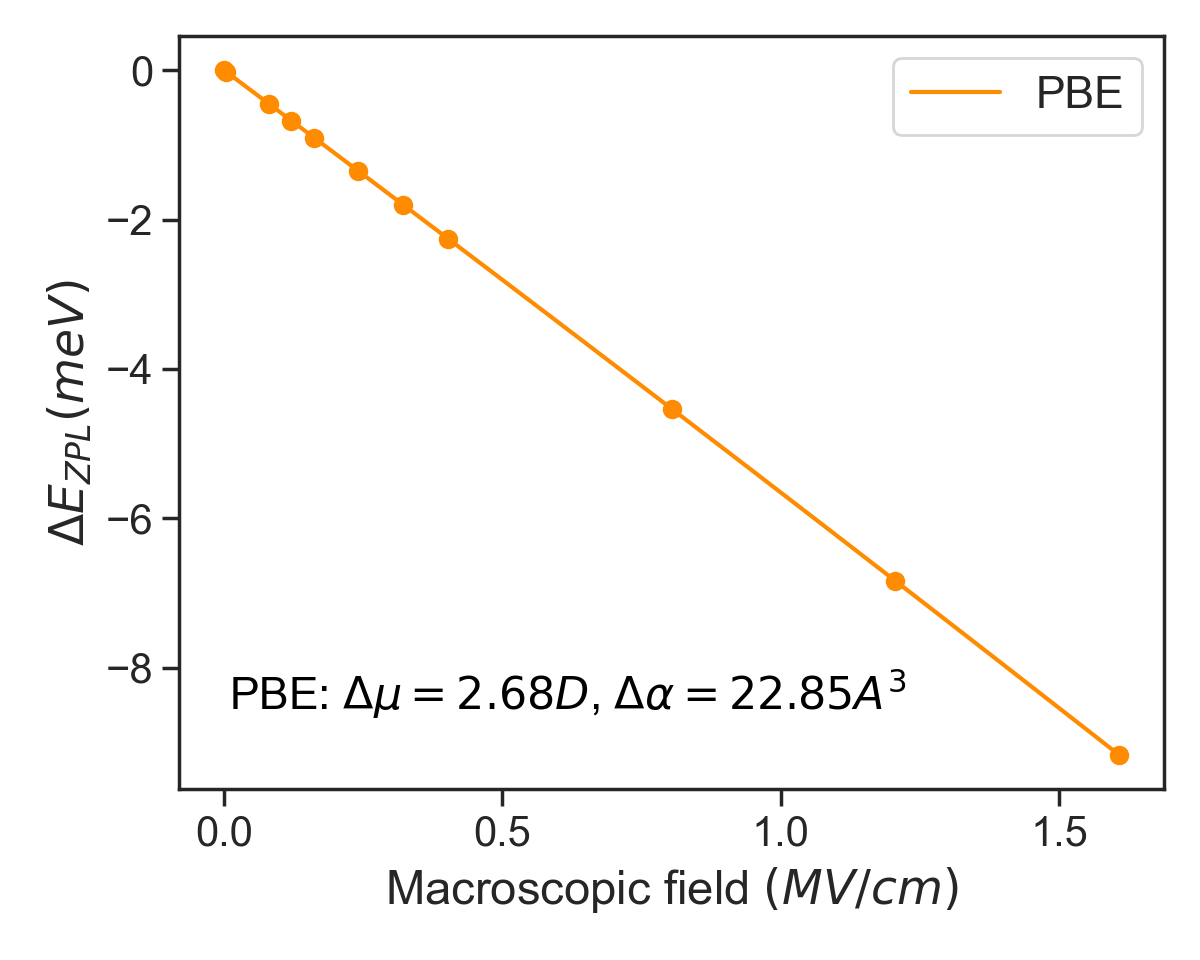}
    \caption{Calculated Stark shift for a 15 \AA\ thick slab with a dopant-defect distance of 9 \AA\ at high electric field}
    \label{fig:convergence_dopant}
\end{figure}

\section{Comparison between bulk and slab calculations}

\subsection{Kohn-Sham eigenvalues}
\begin{figure}
    \centering
    \includegraphics[width=1.0\textwidth]{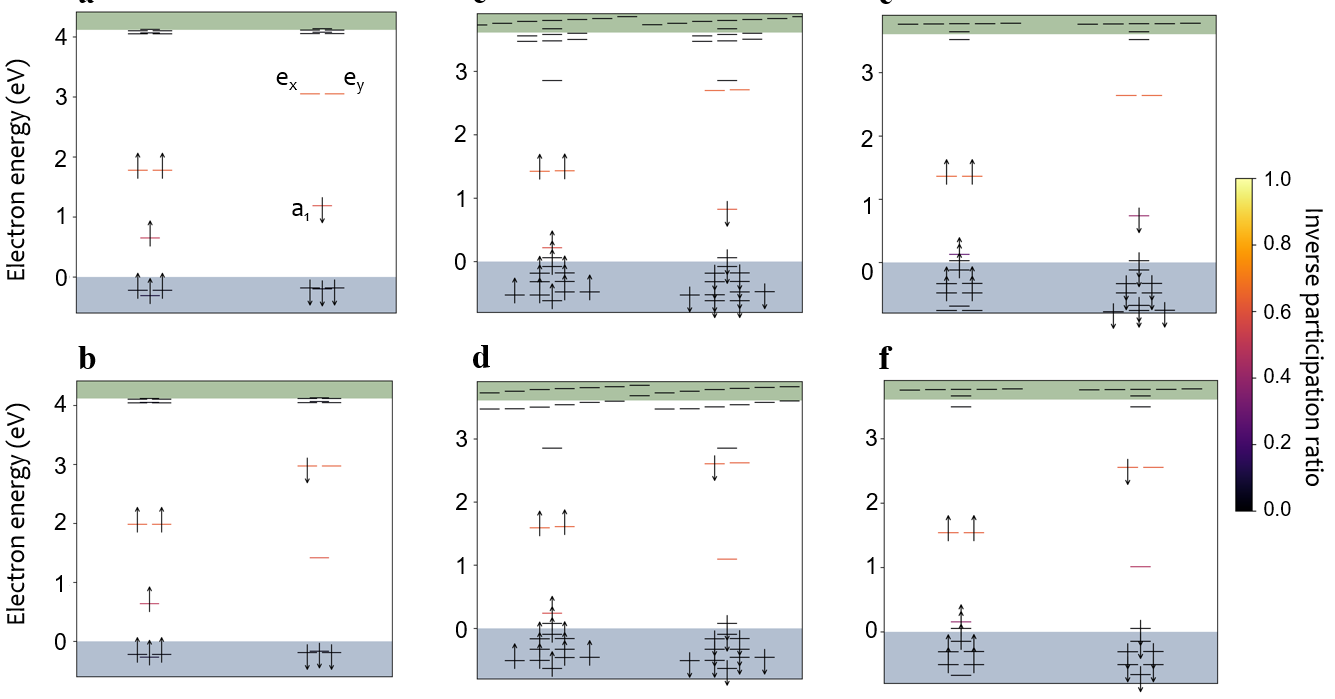}
    \caption{Comparison between the computed Kohn-Sham eigenvalues for bulk and slab calculations (thickness=14.95 \AA) and the ground-state $^3A_2$ and excited-state $^3E$. a) Bulk ground-state, b) excited-state, (c) slab ground-state with a dopant atom (distance $NV-N_{d}$=11.58 \AA) (d) and corresponding excited-state, (e) slab ground-state doped using VCA (side-length=17.69 \AA), (f) and corresponding excited-state.}
    \label{fig:eigenvalues_comparison}
\end{figure}

Then, we compared the Kohn-Sham eigenvalues of our slabs using the two doping schemes with the bulk (Figure \ref{fig:eigenvalues_comparison}). The band gap in the slabs is slightly smaller than in the bulk, decreasing from about 4.1 eV to about 3.6 eV. The most striking difference is the presence of additional levels in the band gap right above the $e_x$ and $e_y$ level when using a dopant atom (see Figure \ref{fig:eigenvalues_comparison} c, d). These levels disappear when the doping is done using the VCA method (see Figure \ref{fig:eigenvalues_comparison}, e, f).

\subsection{Charge density contour plot}
Finally, the charge density contour plots of the $a_1$ and $e_x$ Kohn-Sham levels of the two slabs (doped, VCA) are compared to the bulk.

\begin{figure}[H]
    \centering
    \includegraphics[width=1.0\textwidth]{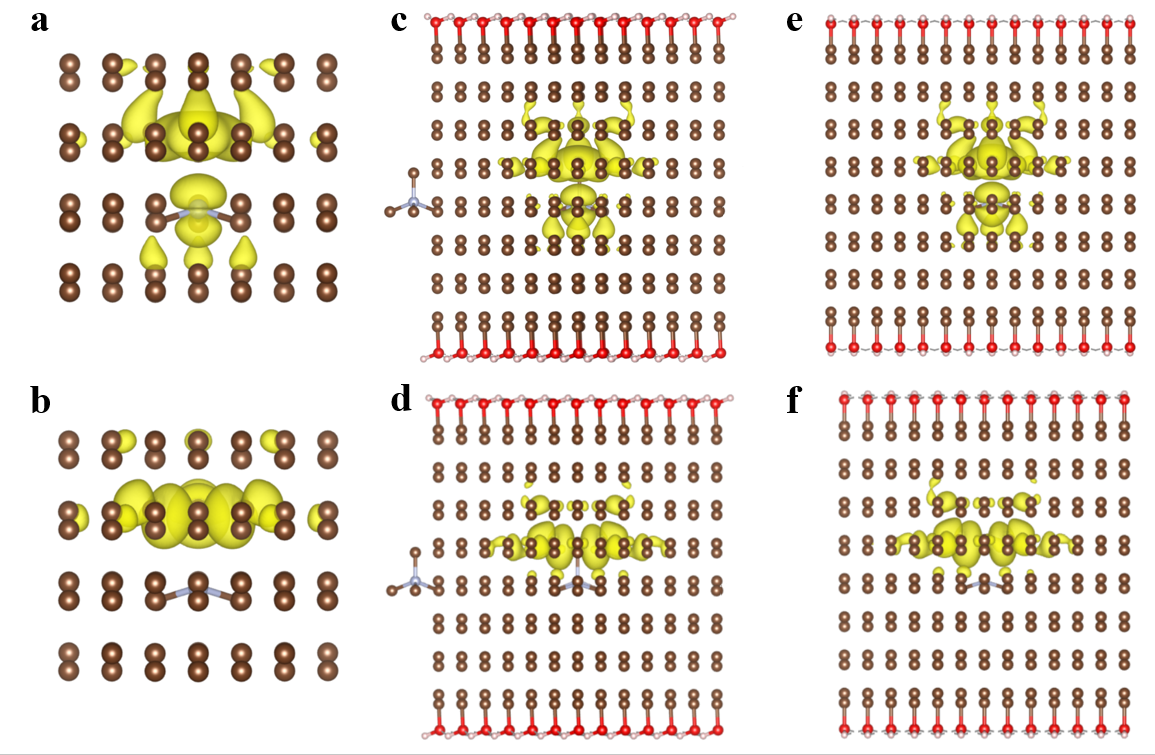}
    \caption{Comparison of charge density contour plots of the $a_1$ (top row) and $e_x$ (bottow row) Kohn-Sham levels of the ground-state for the bulk (a, b), slab with a dopant (c, d) and VCA-doped slab (e, f).}
    \label{fig:charge_density_plot}
\end{figure}

\section{Determination of macroscopic field inside pristine diamond slab}
\begin{figure}[H]
    \centering
    \includegraphics[width=0.85\textwidth]{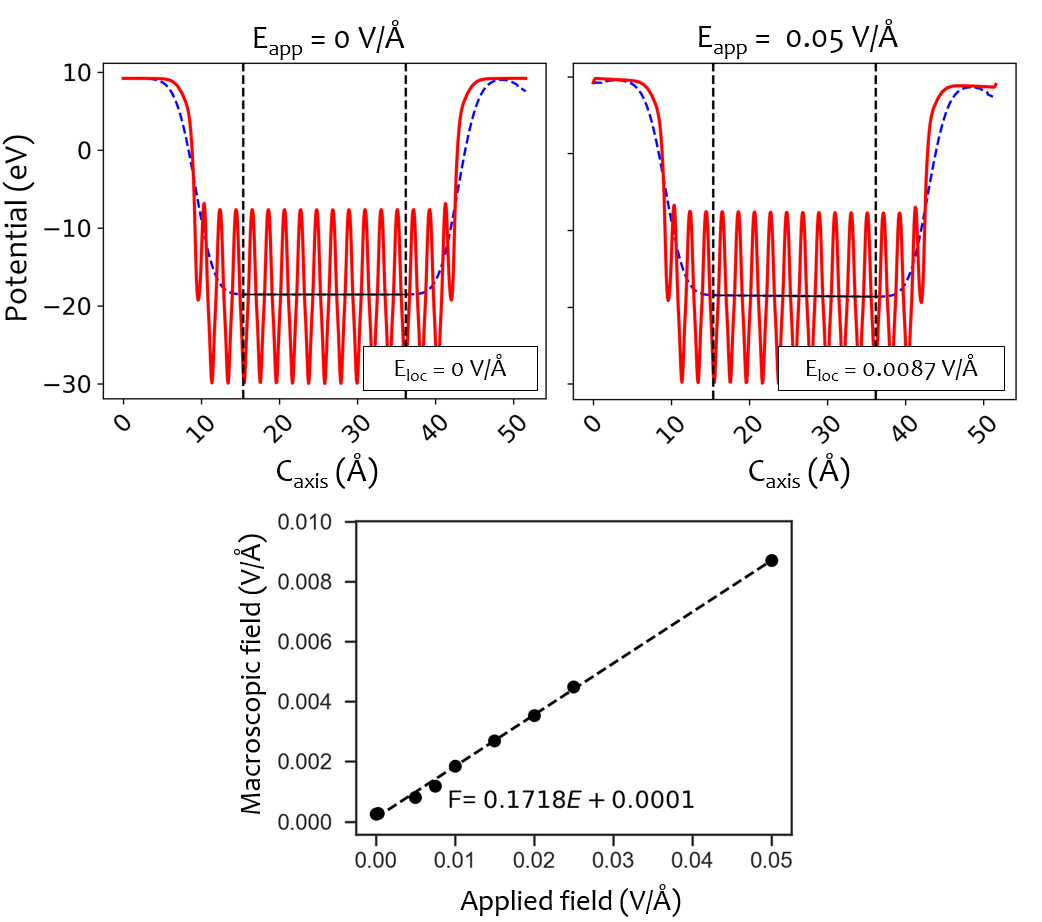}
    \caption{(Top) Electrostatic potential inside a pristine diamond slab terminated by hydroxyl group with and without applied electric field. The red line represents the electrostatic potential, the blue dashed line is the rolling average. The vertical dashed black line shows the points between which the linear fit (black line) was made.
    (Bottom) Relationship between macroscopic field and applied field inside the slab. The inverse of the slope is the dielectric constant of the slab ($\frac{1}{0.1718}=5.82$)}
    \label{fig:macroscopic_field_from_diamond_slab}
\end{figure}

In order to find the relationship between applied and macroscopic electric field, we used a pristine diamond slab terminated by hydroxyl group. Calculations were performed using the Vienna Ab-initio Simulation Package (VASP) with the Perdew-Burke-Ernzerhof (PBE) exchange-correlation functional. The plane-wave cutoff energy was set to 400eV and we sampled the k-space using a 20x20x1 $\Gamma$-centered grid. A vacuum size of 18\AA\ was added in the direction of the applied field to separate periodic images and avoid spurious interactions. The structural relaxation was performed in two steps. First, only the hydroxyl groups and the two outer carbon layers of each side were allowed to move, then the hydroxyl groups were kept fixed and every other site was relaxed. In both cases, relaxation was considered complete once the magnitude of the force on each moving atom was smaller than 0.001eV$/$\AA. No relaxation was allowed when electric fields were applied. The macroscopic electric field was extracted using a rolling average of the local electrostatic potential inside the slab (Figure \ref{fig:macroscopic_field_from_diamond_slab}, top). We then fit the averaged macroscopic field against several values of applied field (Figure \ref{fig:macroscopic_field_from_diamond_slab}, bot). 

\section{Stark shift calculations using Kohn-Sham eigenvalues}
\begin{figure}[H]
    \centering
    \includegraphics[width=1.0\textwidth]{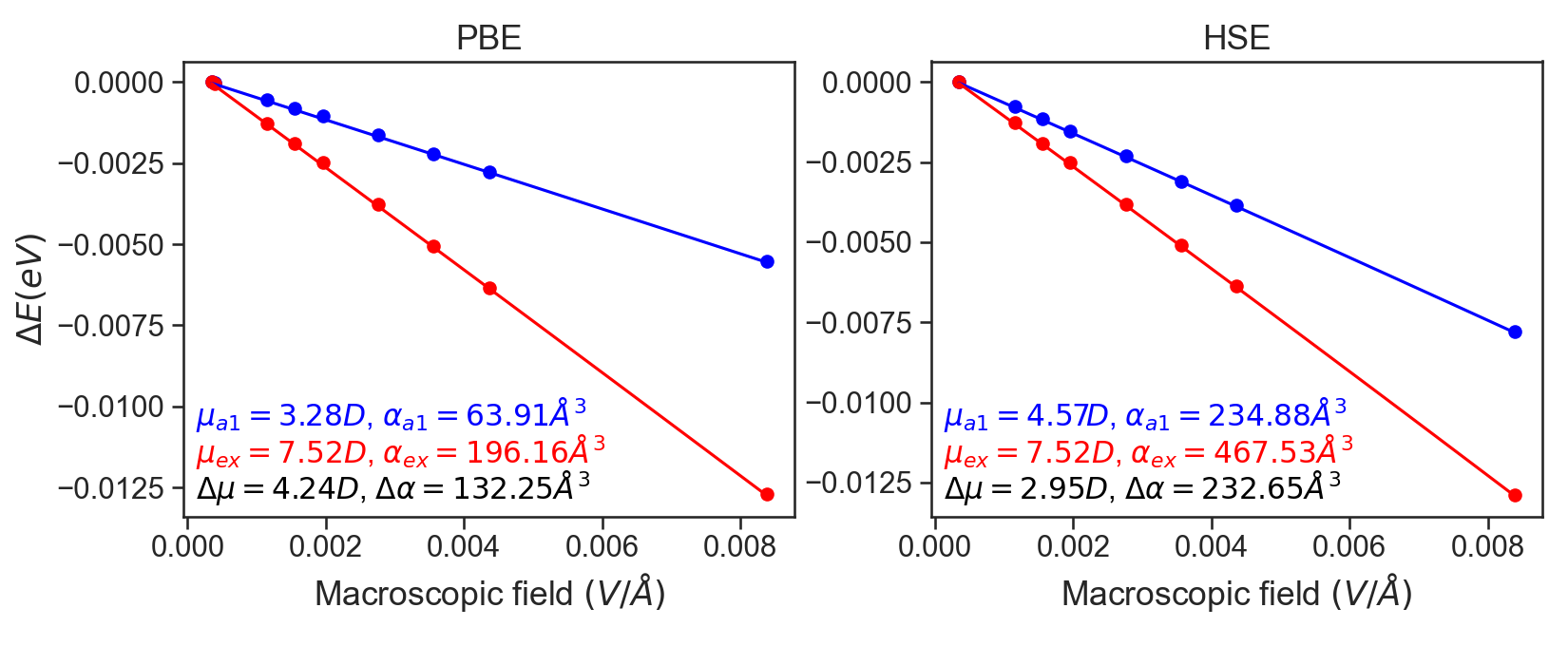}
    \caption{Stark shift using the Kohn-Sham eigenvalues $a_1$ and $e_x$ from the $^3A_2$ ground-state calculation.}
    \label{fig:ks_eigenvalue_shift}
\end{figure}

We explored the possibility to the magnitude of the Stark shift by directly using the Kohn-Sham eigenvalues $a_1$ and $e_x$ from the $^3A_2$ ground-state calculations. This would reduce the computational cost of our methodology as only ground-state calculations are required in this approach. We obtain $\Delta \mu$ = 4.24D, $\Delta \alpha$ = 132.25 $\AA^3$ with PBE and $\Delta \mu$ = 2.95D, $\Delta \alpha$ = 232.65 $\AA^3$ using HSE as shown on Fig. \ref{fig:ks_eigenvalue_shift}.  

\section{Modern theory of polarization calculation}
Due to the implementation issue, directly using the modern theory of polarization to calculate the dipole moment of the excited-state can lead to fallacious results. In this section, we compare the results of two calculation using the modern theory of polarization. We coined the first calculation "direct" modern theory of polarization because it corresponds to the "normal" use of the approach. In the second calculation, or "band-by-band" modern theory of polarization, we explicitly obtain the contribution of each band to the polarization and obtain a dipole moment change in excellent agreement with our slab calculations.

\subsection{Technical details}
All the calculations were performed using the Vienna Ab-initio Simulation Package (VASP version 5.4.4.pl2). We either used the Perdew-Burke-Ernzerhof (PBE) or the hybrid Heyd-Scuseria-Ernzerhof (HSE) exchange-correlation functional. In the latter case, we used 25\% of exact exchange. The defect was placed inside a 4x4x4 supercell (3x3x3 for the HSE calculations) with the nitrogen-vacancy axis oriented in the (111)-direction. The plane-wave cutoff energy was set to 400eV and we used a $\Gamma$-only grid. The energy criterion for the self-consistent-field cycle was set to 1e-6eV.

\subsection{Direct modern theory of polarization}
One of the central result of the modern theory of polarization is that the dipole moment of a periodic system is only defined with respect to a centrosymmetric reference which, by definition, has no dipole moment. Here, the centrosymmetric reference was constructed by shifting the nitrogen atom along the (111)-direction, towards the vacancy, until it sits on an inversion center. This reference belongs to the D3d point group. For the GS, we obtain $\mathbf{\mu}$=(4.37, 4.37, 4.37)D and for the ES, $\mathbf{\mu}$=(1.83, 1.83, 1.83)D. This corresponds to a dipole moment change of 4.43D along (111). 

\begin{figure}[H]
    \centering
    \includegraphics[width=0.85\textwidth]{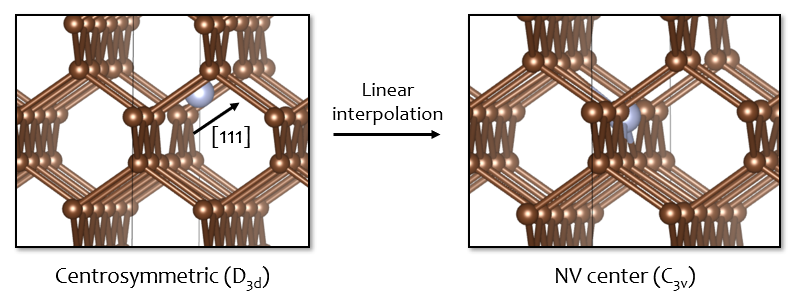}
    \caption{Centrosymmetric reference and actual NV-center structure used for the calculation of the polarization with the Modern theory of polarization. Several snapshot structures are constructed by linear interpolation between the two endpoints to ensure that we stay on the same branch of polarization.}
    \label{fig:berry_illustration}
\end{figure}

\subsection{Band-by-band modern theory of polarization}
To circumvent the problem of the electronic occupation described in the main text, we devised an approach in which the contribution of each band to the dipole moment is individually evaluated. Starting from the calculated wavefunction of the ES, we freeze-in the orbitals and one-electron energies. Then, electrons are artificially added to fill the defect states and the dipole moment is calculated using the modern of polarization formalism for each additional electrons. We obtain the dipole moment of an individual band N by subtracting the calculated dipole moment for the case where N-1 bands are occupied. 

This approach is straightforward to apply in the case of the NV center. We first calculate the dipole moment for the case where only the $a_1$ (1) state is occupied, which also includes all the contribution of the bulk. Compared to the negatively charged NV center, this calculation requires the removal of one electron. We then obtain the contribution of the next state $a_1$ (2) by adding an electron to the system, computing the dipole moment and removing the dipole moment we obtained in the previous case, where only $a_1$ (1) was occupied. Repeating this operation for the $e_x$ and $e_y$ levels, we obtain the dipole moment of each individual state. This allows us to obtain the correct dipole moment of the ES by adding the contribution of each occupied level:

\begin{equation}
    \mu_{ES} = \mu_{bulk} + \mu_{a_1(1)} + \frac{1}{2} (\mu_{e_{x}} + \mu_{e_{y}})
\end{equation}

Using this approach, the calculated dipole moment is now 2.82D using GGA-PBE and 2.23D using HSE (see Table \ref{table:berry_GS} and \ref{table:berry_ES} for the detailed values), far less than what the value obtained by directly using the modern theory of polarization ($\Delta \mu$=4.43D) and in excellent agreement with our slab calculations.

\begin{table}[h!]
    \centering
    \begin{tabular}{lcccc}
    \toprule
    Band &  Occupation   &  $\mu_x$ & $\mu_y$ & $\mu_z$ \\
    \midrule
    Bulk + $a_1$(1)   &   1.0  &   2.554  &   2.554  &   2.554      \\
    $a_1$(2)          &   1.0  &   -0.934  &   -0.934  &   -0.934      \\ 
    $e_x$           &   0.0  &   -1.191  &   -0.014  &   -0.466     \\
    $e_y$           &   0.0  &   0.101 &   -1.077  &  -0.625       \\
    \midrule
    Total GS ($\downarrow$)     &      &  1.620 &   1.620  &  1.620 \\
    \midrule
    \midrule 
    Bulk + $a_1$(1)   &   1.0  &   2.425  &   2.425  &   2.425      \\
    $a_1$(2)         &   0.0  &   -0.756  &   -0.756  &   -0.756      \\ 
    $e_x$           &   0.5  &   -0.334  &   -0.085  &   -1.282      \\
    $e_y$           &   0.5  &   -0.774  &   -1.023  &   0.173       \\
    \midrule
    Total ES ($\downarrow$)       &      &  1.871 &   1.871  &  1.871\\
    \midrule
    $\Delta \mu$ ($\downarrow$)    &   &   0.251   & 0.251 & 0.251     \\
    $\Delta \mu$ ($\uparrow$)    &   &   0.053   & 0.053 & 0.053     \\
    $\Delta \mu$ (ion)    &   &   0.035   & 0.035 & 0.035     \\
    \midrule
    $\Delta \mu$ (total)    &   &   0.339   & 0.339 & 0.339     \\
    \bottomrule
    \end{tabular}
    \caption{Isolated contribution of the bulk and defect states to the dipole moment of the NV center for the GS (top) and ES (bottom) calculated using the PBE functional. Dipole moments are in e\AA. The contribution of the first defect state $a_1$ (1) encompasses the bulk contribution, hence the notation bulk+$a_1$.}
    \label{table:berry_GS}
\end{table}

\begin{table}[h!]
    \centering
    \begin{tabular}{lcccc}
    \toprule
    Band &  Occupation   &  $\mu_x$ & $\mu_y$ & $\mu_z$ \\
    \midrule
    Bulk + $a_1$(1)    &   1.0  &   1.999  &   1.999  &   1.999      \\
    $a_1$(2)         &   1.0  &   0.987  &   0.987  &   0.987      \\ 
    $e_x$           &   0.0  &   0.927  &   1.881  &   0.760      \\
    $e_y$           &   0.0  &   1.515  &   0.560  &  1.682       \\
    \midrule
    Total GS  ($\downarrow$)      &      &  2.986 &   2.986  &  2.986 \\
    \midrule
    \midrule
    Bulk + $a_1$(1)   &   1.0  &   1.943  &   1.943  &   1.943      \\
    $a_1$(2)         &   0.0  &   1.075  &   1.075  &   1.075      \\ 
    $e_x$           &   0.5  &   0.946  &   0.704  &   1.990      \\
    $e_y$           &   0.5  &   1.494  &   1.736  &   0.450       \\
    \midrule
    Total ES  ($\downarrow$)        &       &  3.163  &  3.163   &  3.163  \\
    \midrule
    $\Delta \mu$ ($\downarrow$)    &   &   0.177   & 0.177 & 0.177  \\
    $\Delta \mu$ ($\uparrow$)    &   &   0.058   & 0.058 & 0.058     \\
    $\Delta \mu$ (ion)    &   &   0.032   & 0.032 & 0.032     \\
    \midrule
    $\Delta \mu$ (total)    &   &   0.267   & 0.267 & 0.267     \\
    \end{tabular}
    \caption{Isolated contribution of the bulk and defect states to the dipole moment of the NV center for the GS (top) and ES (bottom) calculated using the HSE functional. Dipole moments are in e\AA. The contribution of the first defect state $a_1$ (1) encompasses the bulk contribution, hence the notation bulk+$a_1$.}
    \label{table:berry_ES}
\end{table}

The code used to generate the inputs and process the results of these calculations is available on Github. 

\bibliography{biblio}